\documentclass{article}
\usepackage{spconf,amsmath,graphicx}

\usepackage{enumitem}
\usepackage{algorithm}
\usepackage{algorithmic}
\usepackage{makecell}
\usepackage{multirow}
\setlist{nosep, leftmargin=14pt}
\usepackage{tikz}
\usepackage{calc}
\def\checkmark{\tikz\fill[scale=0.3](0,.35) -- (.25,0) -- (1,.7) -- (.25,.15) -- cycle;} 
\usepackage{mwe} 

\title{Deep Supervision by Gaussian Pseudo-label-based Morphological Attention for Abdominal Aorta Segmentation in Non-Contrast CTs}
\name{Author(s) Name(s)\thanks{Some author footnote.}}
\name{Qixiang Ma\textsuperscript{*,1,2}\thanks{*Correspondence: qixiang.ma@etudiant.univ-rennes1.fr}, Antoine Lucas\textsuperscript{1,2}, Adrien Kaladji\textsuperscript{1,2}, Pascal Haigron\textsuperscript{1,2}}
\address{\textsuperscript{1}Univ Rennes, CHU Rennes, Inserm, LTSI – UMR 1099, Rennes, France\\
\textsuperscript{2}Centre de Recherche en Information Biomédicale Sino-français (CRIBs), Univ Rennes, Inserm, \\
Southeast University, Rennes, France, Nanjing, China}

\begin{document}
%
\maketitle
\begin{abstract}
The segmentation of the abdominal aorta in non-contrast CT images is a non-trivial task for computer-assisted endovascular navigation, particularly in scenarios where contrast agents are unsuitable. While state-of-the-art deep learning segmentation models have been proposed recently for this task, they are trained on manually annotated strong labels. However, the inherent ambiguity in the boundary of the aorta in non-contrast CT may undermine the reliability of strong labels, leading to potential overfitting risks. This paper introduces a Gaussian-based pseudo label, integrated into conventional deep learning models through deep supervision, to achieve Morphological Attention (MA) enhancement. As the Gaussian pseudo label retains the morphological features of the aorta without explicitly representing its boundary distribution, we suggest that it preserves aortic morphology during training while mitigating the negative impact of ambiguous boundaries, reducing the risk of overfitting. It is introduced in various 2D/3D deep learning models and validated on our local data set of 30 non-contrast CT volumes comprising 5749 CT slices. The results underscore the effectiveness of MA in preserving the morphological characteristics of the aorta and addressing overfitting concerns, thereby enhancing the performance of the models.
\end{abstract}
\section{Introduction}
\label{sec:intro}
Abdominal aortic aneurysm (AAA) is a prevalent and life-threatening vascular disease \cite{robinson2013derivation}. Clinical intervention is often imperative for patients to mitigate the risk of aortic aneurysm rupture. Endovascular aneurysm repair (EVAR) has emerged as a common interventional procedure, offering a risk-averse alternative to open surgery. EVAR procedures are typically guided by contrast-enhanced computed tomography angiography (CTA). The use of contrast agents facilitates the visualization of the aortic lumen in high contrast within the CT images, allowing for the segmentation of the abdominal aortic region and its projection onto 2D intra-operative images, thereby guiding the EVAR process \cite{kaladji2015safety}. However, in specific scenarios where the use of contrast agents is contraindicated, such as in patients with renal impairment susceptible to contrast-induced complications \cite{walsh2008renal, mehran2006contrast}, alternative approaches become essential. Consequently, in situations where contrast agents are unsuitable, non-contrast CT serves as a viable means for guiding intervention.

Kaladji et al. \cite{kaladji2015safety} have demonstrated the safety and accuracy of non-contrast CT-guided EVAR. However, their strategy to obtain aorta segmentation from non-contrast CT involved manual segmentation which is labor-intensive. In recent years, the advent of Deep Learning (DL) techniques has made fully automated segmentation of medical images feasible. State-of-the-art DL models, exemplified by Convolutional Neural Networks (CNN) \cite{ronneberger2015u,oktay2018attention,milletari2016v} and Transformer-based \cite{chen2021transunet,wang2021transbts} models, have emerged as popular choices for the abdominal aortic segmentation in non-contrast CTs. Previous studies have progressively delved into applying DL models for abdominal aortic segmentation in non-contrast CTs. Researchers such as Lu et al. \cite{lu2019deepaaa} and Chandrashekar et al. \cite{chandrashekar2020deep} have respectively incorporated 3D DL models into the segmentation process of non-contrast CTs. Lu et al. employed a deformed 3D U-net, while Chandrashekar et al. integrated common DL mechanisms such as attention and cascade methods into their approach. Ma et al. \cite{ma2023deep} proposed a 2D-3D feature fusion approach to address the challenge of low contrast in abdominal aortic segmentation in non-contrast CTs.

Despite the commendable performance exhibited by the aforementioned DL methods in the segmentation of the abdominal aorta in non-contrast CTs, they rely on manually provided strong labels for training and aim to achieve optimal generalization. However, owing to the inherent characteristics of non-contrast CT, the boundaries between the aorta and surrounding tissues are often ambiguous. This ambiguity can lead to divergent definitions of the aortic boundary among different observers or even within the same observer for a given CT scan, resulting in intra-observer/inter-observer variability. If a DL model is trained solely to fit the boundary defined ambiguously, there is a potential risk of overfitting, diminishing the model's generalization capabilities on other data. 

In classical machine learning classification problems, an effective classifier aims to achieve an "appropriate" fit on the training set rather than an "over" fit, as overly fitting the training data may diminish its generalization performance on other datasets. Methods to prevent overfitting involve regularization terms to penalize rapid convergence of the loss function and make it fit a general boundary. Motivated by this wisdom, we hypothesize that the segmentation of the abdominal aorta in non-contrast CTs can follow a similar principle. Specifically, we introduce a smooth structure resembling the morphology of the aorta as a "regularization term" to conventional DL models. Considering the elliptical approximation of the aorta in CT slices, we define this structure as a two-dimensional Gaussian distribution representing the best-fitting ellipse for the aorta. This smooth structure retains the aortic morphology in CT slices without explicit boundary, allowing the model to mitigate the risk of overfitting to ambiguous boundaries in strong labels. Inspired by deep supervision \cite{lee2015deeply}, we employ this smooth structure as a pseudo label for the sub-output of the decoder, correcting only the intermediate features of the model. We term this process the "morphological attention (MA)" mechanism.

In this study, we propose a morphological attention (MA) mechanism based on Gaussian pseudo labels to preserve the morphology of the aorta in non-contrast CT slices and mitigate the potential risk of overfitting. We first describe the proposed mechanism which is applicable to any segmentation model based on an encoder-decoder architecture through a plug-and-play way. Then we conduct the experimental validation of this mechanism on our local non-contrast CT data to show its feasibility and efficacy.
\section{METHODOLOGY}
\label{sec:method}
\subsection{Overall Framework}
\begin{figure}[!t]
\begin{minipage}[c]{1.0\linewidth}
\includegraphics[width=\textwidth]{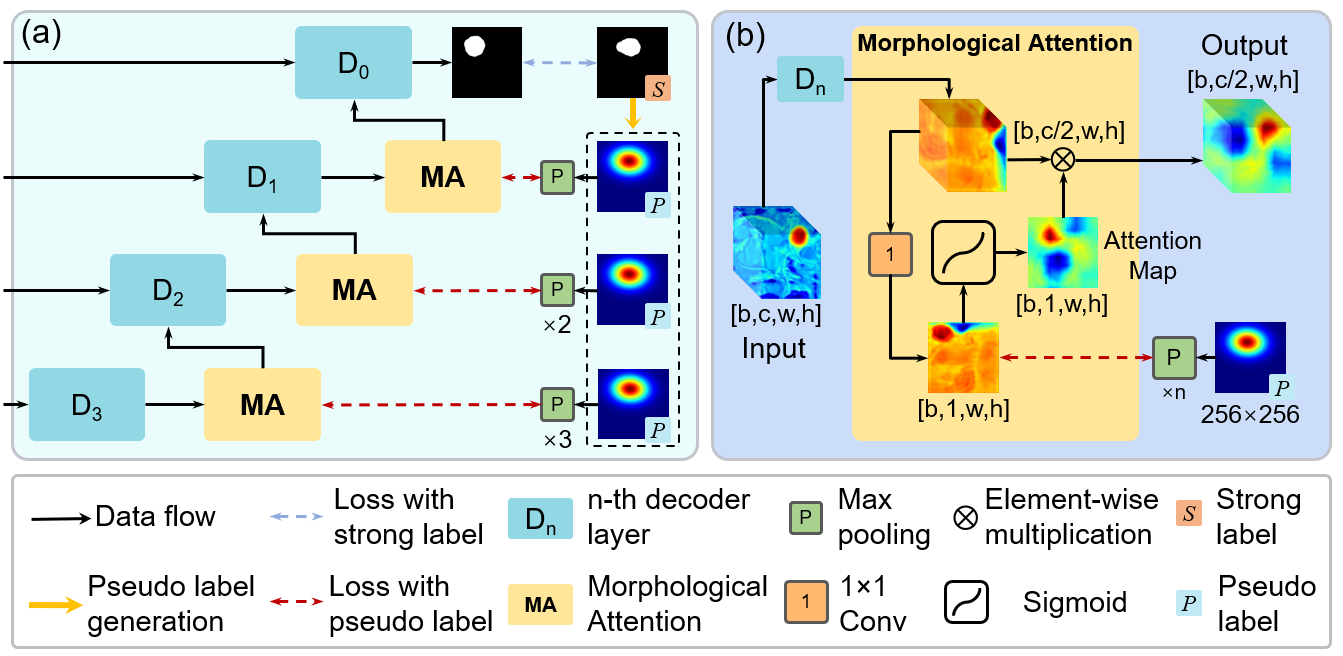}
\end{minipage}
\caption{(a) The general decoder architecture of DL models involving Morphological Attention. (b) Detailed mechanism of Morphological Attention.}\label{fig:f1}
\end{figure}

Fig.~\ref{fig:f1} illustrates a general decoder architecture of the DL-based segmentation model and the application of Morphological Attention. Specifically, Fig.~\ref{fig:f1}(a) shows the morphological attention enhancement applied to the sub-outputs of each decoder layer through pseudo labels generated at different scales, where the pseudo labels are generated from their original strong labels. The strong labels optimize the final network output as in conventional DL models. The optimization of each layer is assigned a specific weight. Therefore, an end-to-end training with a hybrid of strong labels and pseudo labels is achieved. The detailed mechanism of Morphological Attention (MA) is depicted in the yellow region of Fig.~\ref{fig:f1}(b). In this process, the feature map from the $n$-th layer undergoes dimension reduction through a 1$\times$1 convolution within the MA module. It is then refined by the pseudo label of the same scale (red dashed line) through a loss function and activated by a sigmoid function, which generates an attention map that interacts with the original features through element-wise multiplication, producing attention-enhanced features for the next layer. 

\begin{figure}[!t]
\begin{minipage}[c]{1.0\linewidth}
\includegraphics[width=\textwidth]{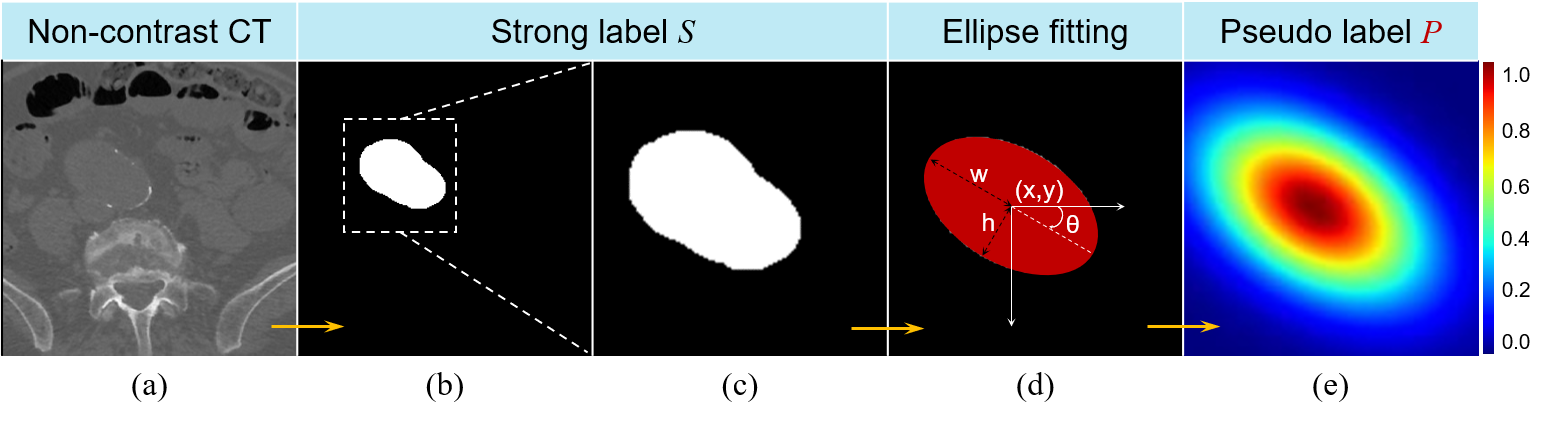}
\end{minipage}
\caption{The produce of generation of pseudo labels $P$ from strong labels $S$.}\label{fig:f2}
\end{figure}

Since the pseudo label only reflects the morphological features resembling an elliptical form during the refinement of intermediate features without explicitly representing boundary information, we suggest that this mechanism enables the network to focus more on the "general" morphology of the aorta in CT slices rather than overfitting to specific boundary distributions.
\subsection{Pseudo Label Generation}
Fig.~\ref{fig:f2} illustrates the process of generating pseudo labels. Specifically, the parameters (central point $(x,y)$, the major and minor axes $w$ and $h$, the rotation angle $\theta$) of the best-fitting ellipse of the strong label $S$ are obtained through an ellipse-fitting algorithm. These parameters are then utilized to generate a 2D heatmap corresponding to the strong label. The amplitude of this heatmap follows a Gaussian distribution, and it serves as the pseudo label $P$. The details of pseudo-label generation are exhibited in Algorithm 1.

Given a strong label $S$, the process begins by extracting its boundary $S'$. Subsequently, an ellipse-fitting algorithm is applied to obtain the parameters of the best-fitting ellipse of $S'$. The five parameters are the central point $(x,y)$, the major and minor axes $w$ and $h$, and the rotation angle $\theta$. The ellipse fitting algorithm employed in this study is the numerically stable least squares algorithm \cite{halir1998numerically}. Then, two orthogonal uniform distributions are initialized, followed by centering rotation according to the parameters of the fitted ellipse. Each distribution undergoes Gaussian transformation independently, resulting in marginal distributions with amplitudes ranging from 0 to 1. These distributions are then element-wise multiplied to obtain their joint distribution, which serves as the pseudo label $P$ corresponding to $S$.
\begin{algorithm}
\label{alg:Labertian}
\caption{Pseudo Label Generation.}
\begin{algorithmic}[1]
\REQUIRE Strong label $S$.
\ENSURE Pseudo label $P$. \\
\STATE $S'$ = BoundaryExtraction($S$)
$//$ \emph{Boundary extraction to obtain the boundary of $S$.}\\
\STATE $x,y,w,h,\theta$ = EllipseFitting($S'$)
$//$ \emph{Obtaining the parameters of the best-fitting ellipse of $S'$}\\
\STATE 
${U_x} = \left( {\begin{array}{*{20}{c}}
{0}&{\cdots}& {255}\\
 {\vdots}&{\ddots}&{\vdots}\\
{0}&{\cdots}& {255}
\end{array}} \right)$, ${U_y} = \left( {\begin{array}{*{20}{c}}
0&\cdots& 0\\
{\vdots}&{\ddots}&{\vdots}\\
{255}&{\cdots}&{255}
\end{array}} \right)$
$//$ \emph{Initializing two discrete uniform distributions, Assuming that the spatial size is $256\times 256$}
\STATE ${M_x} = {U_x} - x{\mbox{, }}{M_y} = {U_y} - y$
$//$ \emph{Localizing the central position}
\STATE ${P_x} = {M_x}\cos \theta  + {M_y}\sin \theta {\mbox{, }}{P_y} = {M_y}\cos \theta  + {M_x}\sin \theta$
$//$ \emph{Rotating the distributions with the rotation angle $\theta$}
\STATE 
${f_{Px}}(t) = \frac{1}{{{w}\sqrt {2\pi } }}{e^{ - \frac{{{{(t - x)}^2}}}{{2w^2}}}}{\mbox{, }}{f_{Py}}(t) = \frac{1}{{{h}\sqrt {2\pi } }}{e^{ - \frac{{{{(t - y)}^2}}}{{2h^2}}}}$
$//$ \emph{Gaussianization (1). Initialization of Gaussian Probability Density Functions (PDFs)}
\STATE 
${F_x} = \frac{{{f_{Px}}(t)}}{{{f_{Px}}(t = x)}} = {e^{ - \frac{{{{(t - x)}^2}}}{{2w^2}}}}{\mbox{, }}{F_y} = \frac{{{f_{Py}}(t)}}{{{f_{Py}}(t = x)}} = {e^{ - \frac{{{{(t - y)}^2}}}{{2h^2}}}}$
$//$ \emph{Gaussianization (2). 0-1 normalization for the two PDFs.}
\STATE
${F_x} = {e^{ - \frac{{{{P_x}^2}}}{{2w^2}}}}{\mbox{, }}{F_y} = {e^{ - \frac{{{{P_y}^2}}}{{2h^2}}}}$
$//$ \emph{Let $t-x=P_x$, $t-y=P_y$}
\STATE
$P = {F_x} \otimes {F_y}$
$//$ \emph{Integrating the two orthogonal distributions into a joint distribution}
\RETURN $P$ 
\end{algorithmic}
\end{algorithm}
\subsection{Loss Functions}
Let $x_n$ be the output of the $n$-th layer, $S$ and $P$ be the strong and pseudo label, respectively. The loss function of the entire model is defined as
\begin{equation}
\begin{array}{l}
\mathcal{L} = {w_0}(Dice({x_0},S) + BCE({x_0},S))\\
 + \sum\nolimits_{i = 1}^{n - 1} {{w_n}{L_\infty }(|{x_n} - Poolin{g_n}(P)|)},
\end{array}
\end{equation}
where $Dice( \cdot )$ and $BCE( \cdot )$ represent the Dice loss \cite{milletari2016v} and Binary Cross Entropy loss, respectively. $L_\infty$ means the L-Infinity Norm outperforming the $L_1$ and $L_2$ Norm in our practice, where
\begin{equation}
{L_\infty }({\bf{X}}) = |{\bf{X}}{|_\infty } = \mathop {\max }\limits_i |{x_i}|.
\end{equation}
The $Pooling_n$ means $n\times$  Max Pooling processes are applied. $w_n$ is the weight of the loss for $n$-th layer, where
\begin{equation}
{w_n} = {{\frac{1}{{{2^n}}}} \mathord{\left/
 {\vphantom {{\frac{1}{{{2^n}}}} {\sum\nolimits_{i = 0}^{n - 1} {\frac{1}{{{2^n}}}} }}} \right.
 \kern-\nulldelimiterspace} {\sum\nolimits_{i = 0}^{n - 1} {\frac{1}{{{2^n}}}} }}.
\end{equation}
\section{EXPERIMENTS}
\label{sec:ex}
\subsection{Data set and Implementation Details}
We validated our approach on a local dataset comprising 5749 slices of non-contrast CTs from 30 patients diagnosed with abdominal aortic aneurysms. The original data consisted of DICOM files with a size of 512$\times$512 pixels and slice thickness ranging from 0.625 to 5 mm. Two experts, $A$ and $B$, independently labeled the data, where $A$ labeled it twice at a ten-day interval. These annotations were used for an intra- and inter-observer variability analysis. According to the Dice score, the intra- and inter-observer variabilities were 96.6\% and 96.1\%, respectively. Such consistencies show the reliability of the labels. We selected the annotations from the first session of A and extracted a 256$\times$256 region of interest (RoI) as the final experimental data. A 3-fold cross-validation was performed, each containing 10 volumes for testing. The remaining 20 volumes were used for training (15) and validation (5).

The DL models underwent training for 300 epochs, with the best weights on the validation set used for testing. The adam optimization \cite{kingma2014adam} was employed with an initial learning rate of $r=0.001$, linearly decreasing by a ratio of $1-r \times n$ for each epoch, with $\beta_1=0.9$, $\beta_2=0.999$, weight decay was $1\times 10^{-10}$ during training. The 2D and 3D models had an input size of 256$\times$256 (with a batch size 16) and 16$\times$256$\times$256, respectively. They were normalized to zero mean and unit variance before being fed into the model. During training, data augmentation was applied to the training data, including random horizontal and vertical flips, rotation with an angle varying in $[-\pi, \pi]$, perspective distortion with a scale of 0.2, and a Gaussian Blur with a kernel size of $3\time 3$.
\subsection{Results}
\subsubsection{Ablation Study}
We initially investigated the impact of general Deep Supervision (DS) and Morphological Attention (MA) on enhancing the performance of the DL model. Table~\ref{tab:ex1} presents the performance comparison between the standard U-net and its counterparts incorporating general DS and MA, where the general DS involves multi-scaled strong labels. It is observed that the U-net augmented with MA exhibits more substantial improvements in both Dice Similarity Coefficient (DSC) and Hausdorff distance (HD) compared to the version with general Deep Supervision, where the DS reduces the Sensitivity (SEN) of the original performance. It suggests that MA's enhancement in model performance is not solely attributable to Deep Supervision but is also influenced by the attention mechanism based on pseudo labels.

\begin{figure}[!t]
\begin{minipage}[c]{1.0\linewidth}
\includegraphics[width=\textwidth]{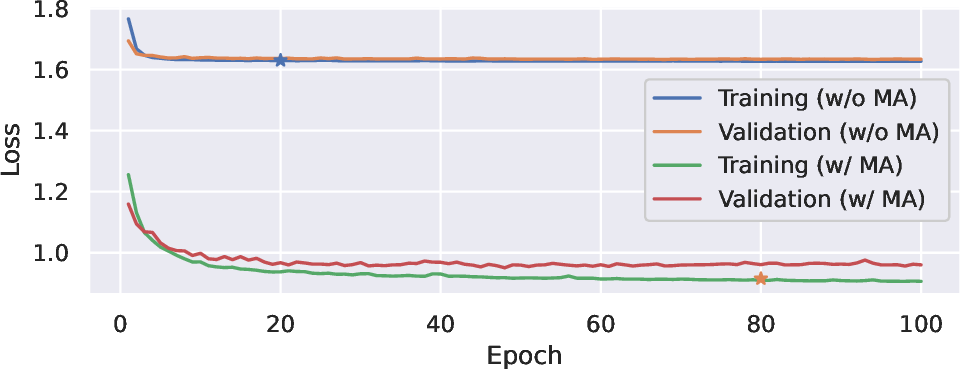}
\end{minipage}
\caption{Convergence of loss of training and validation data of U-net with and without MA. The asterisk (*) denotes the epochs when convergence approaches saturation.}\label{fig:f3}
\end{figure}

\begin{table}[h]
\setlength{\tabcolsep}{0.7mm}
\caption{Performance comparison of the original U-net, the DS and MA enhancement.}
\begin{center}
\begin{tabular}{cccccc}
\hline
Model                            &DSC (\%)$\uparrow $                                &SEN (\%)$\uparrow $                         &HD$\downarrow $\\
\hline
U-net \cite{ronneberger2015u}    &85.7\scriptsize$\pm$1.1                            &88.7\scriptsize$\pm$2.7                     &10.40\scriptsize$\pm$0.99\\
U-net+DS                         &86.7{\scriptsize$\pm$1.3} \textbf{($\uparrow $1.0)}&88.0{\scriptsize$\pm$3.0} ($\downarrow $0.7)&8.72{\scriptsize$\pm$2.60} \textbf{($\downarrow $1.68)}\\
U-net+\textbf{MA}                &88.2{\scriptsize$\pm$1.1} \textbf{($\uparrow $2.5)}&88.7{\scriptsize$\pm$2.3}                   &6.89{\scriptsize$\pm$0.69} \textbf{($\downarrow $3.51)}\\
\hline 
\end{tabular}
\end{center}
\label{tab:ex1}
\end{table}
Figure~\ref{fig:f3} illustrates the convergence of loss of training and validation data of U-net with and without MA. Compared to the original model (converging at epoch=20), the model involving MA converges more slowly (at epoch=80). Considering that the best weights on the validation set will be used for testing, we suggest that premature convergence of the model might impact its generalization on the test set. The pseudo-labels-based MA serves as a regularization term, mitigating premature fitting issues and reducing the risk of overfitting.
\subsubsection{Performance of state-of-the-art enhanced by MA}
\begin{figure}[!t]
\begin{minipage}[c]{1.0\linewidth}
\includegraphics[width=\textwidth]{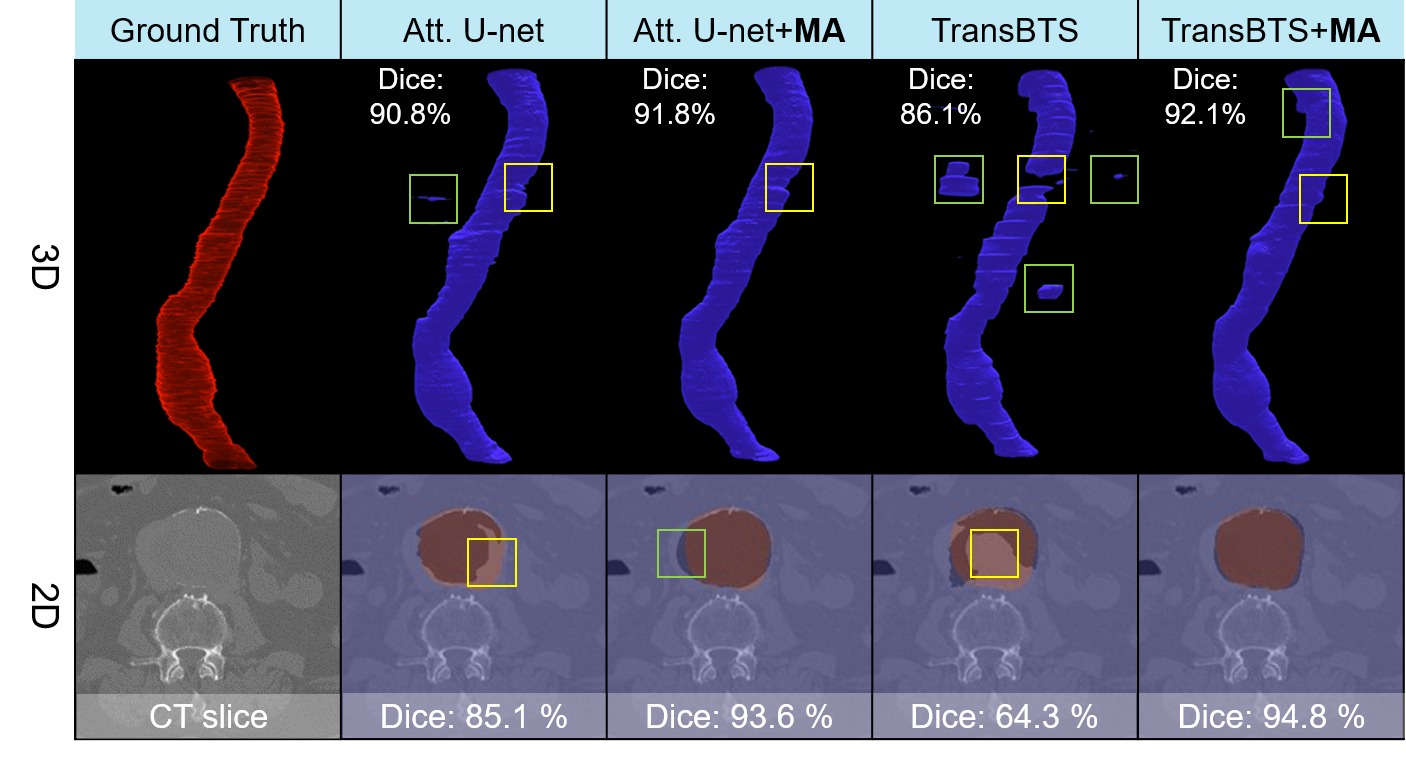}
\end{minipage}
\caption{3D/2D results from DL models with and without abdominal aorta (MA) improvement. In 2D, red and blue regions indicate ground truth and predictions, while purple represents overlay areas. Green and yellow dotted boxes are False Positives (FPs) and False Negatives (FNs).}\label{fig:f4}
\end{figure}

Table~\ref{tab:ex2} presents the performance of state-of-the-art models equipped with MA, including 2D/3D models based on CNN and Transformer architectures. The models enhanced with MA outperform their original models on both metrics. Particularly noteworthy is the significant reduction in HD for both 3D models, indicating better preservation of the aortic morphology. This improvement is attributed to the pseudo labels, enabling the model to learn a more "general" aortic morphology without overfitting to complex boundaries. The qualitative results in Figure~\ref{fig:f4} substantiate this viewpoint, demonstrating that both 2D and 3D models augmented with MA exhibit improved performance. It is observed that MA effectively reduces False Positives and preserves the morphological characteristics of the aorta in CT slices. This phenomenon is particularly pronounced in the 3D model (TransBTS).
\begin{table}[h]
\setlength{\tabcolsep}{1.2mm}
\caption{Performance comparison of the state-of-the-art 2D/3D models and the enhancement of MA.}
   \begin{center}
  \begin{tabular}{ccccc}
\hline
    &Model   &MA &DSC (\%)$\uparrow $&HD$\downarrow $\\
\hline
   \multirow{4}{*}{2D}&\multirow{2}{*}{\makecell{Attention\\U-Net \cite{oktay2018attention}}}&~&86.5\scriptsize$\pm$2.1&8.45\scriptsize$\pm$1.90\\
   ~& ~&\checkmark&88.4{\scriptsize$\pm$1.6} \textbf{($\uparrow $1.9)}&6.45{\scriptsize$\pm$0.94} \textbf{($\downarrow $2.00)}\\
   \cline{2-5}
   ~&\multirow{2}{*}{\makecell{Trans-\\UNet \cite{chen2021transunet}}}&~&87.1\scriptsize$\pm$2.3&8.46\scriptsize$\pm$0.62\\
   ~& ~&\checkmark&88.6{\scriptsize$\pm$1.3} \textbf{($\uparrow $1.5)}&5.32{\scriptsize$\pm$0.41} \textbf{($\downarrow $3.14)}\\
\hline
   \multirow{4}{*}{3D}&\multirow{2}{*}{\makecell{V-Net \cite{milletari2016v}}}&~&84.3\scriptsize$\pm$1.6&17.93\scriptsize$\pm$7.39\\
   ~& ~&\checkmark&85.0{\scriptsize$\pm$1.6} \textbf{($\uparrow $0.7)}&10.62{\scriptsize$\pm$1.46} \textbf{($\downarrow $7.31)}\\
   \cline{2-5}
   ~&\multirow{2}{*}{\makecell{Trans-\\BTS \cite{wang2021transbts}}}&~&85.1\scriptsize$\pm$0.5&18.85\scriptsize$\pm$6.97\\
   ~& ~&\checkmark&87.1{\scriptsize$\pm$2.4} \textbf{($\uparrow $2.0)}&8.41{\scriptsize$\pm$1.89} \textbf{($\downarrow $10.44)}\\                                              
\hline
  \end{tabular}
  \end{center}
  \label{tab:ex2}
\end{table}
\vspace{-0.8cm} 

\section{Conclusion}\label{sec:con}
We proposed a Morphological Attention method based on Gaussian pseudo labels, integrated into a conventional DL model through Deep Supervision. We specially highlighted its dual roles: 1) addressing the potential overfitting issues arising from the inherent ambiguity of boundary in the strong labels of abdominal aorta segmentation in non-contrast CTs and 2) effectively preserving the morphology of the aorta. We attribute this success to the pseudo labels. The results showed that they not only capture the general morphology of the aorta but also serve as a "regularization term" during training to mitigate the risk of overfitting. They also demonstrate superiority of MA over conventional Deep Supervision and its ability to enhance network performance in a plug-and-play manner. The limitation of this study is that pseudo labels have not been validated on more complex vascular structures. Therefore, our future research aims to extend the application of pseudo labels to other blood vessels and incorporate them into network training in a more flexible manner, particularly in the context of weak supervision.

\section{Compliance with ethical standards}
\label{sec:ethics}
This retrospective study was performed in line with the principles of the Declaration of Helsinki. Approval was granted by our IRB.

\section{Acknowledgments}
\label{sec:acknowledgments}
This study was partially supported by the French National Research Agency (ANR) in the framework of the Investissement d’Avenir Program through Labex CAMI (ANR-11- LABX-0004). The first author is grateful for the support of China Scholarship Council (CSC Grant No. 201906090389)

\bibliographystyle{IEEEbib}
\bibliography{strings,refs}

\begin{thebibliography}{10}

\bibitem{robinson2013derivation}
William~P Robinson, Andres Schanzer, YouFu Li, Philip~P Goodney, Brian~W Nolan,
  Mohammad~H Eslami, Jack~L Cronenwett, and Louis~M Messina,
\newblock ``Derivation and validation of a practical risk score for prediction
  of mortality after open repair of ruptured abdominal aortic aneurysms in a us
  regional cohort and comparison to existing scoring systems,''
\newblock {\em Journal of vascular surgery}, vol. 57, no. 2, pp. 354--361,
  2013.

\bibitem{kaladji2015safety}
Adrien Kaladji, Aur{\'e}lien Dumenil, Guillaume Mah{\'e}, Miguel Castro, Alain
  Cardon, Antoine Lucas, and Pascal Haigron,
\newblock ``Safety and accuracy of endovascular aneurysm repair without
  pre-operative and intra-operative contrast agent,''
\newblock {\em European Journal of Vascular and Endovascular Surgery}, vol. 49,
  no. 3, pp. 255--261, 2015.

\bibitem{walsh2008renal}
Stewart~R Walsh, Tjun~Y Tang, and Jonathan~R Boyle,
\newblock ``Renal consequences of endovascular abdominal aortic aneurysm
  repair,''
\newblock {\em Journal of Endovascular Therapy}, vol. 15, no. 1, pp. 73--82,
  2008.

\bibitem{mehran2006contrast}
R~Mehran and E~Nikolsky,
\newblock ``Contrast-induced nephropathy: definition, epidemiology, and
  patients at risk,''
\newblock {\em Kidney international}, vol. 69, pp. S11--S15, 2006.

\bibitem{ronneberger2015u}
Olaf Ronneberger, Philipp Fischer, and Thomas Brox,
\newblock ``U-net: Convolutional networks for biomedical image segmentation,''
\newblock in {\em Medical Image Computing and Computer-Assisted
  Intervention--MICCAI 2015: 18th International Conference, Munich, Germany,
  October 5-9, 2015, Proceedings, Part III 18}. Springer, 2015, pp. 234--241.

\bibitem{oktay2018attention}
Ozan Oktay, Jo~Schlemper, Loic~Le Folgoc, Matthew Lee, Mattias Heinrich,
  Kazunari Misawa, Kensaku Mori, Steven McDonagh, Nils~Y Hammerla, Bernhard
  Kainz, et~al.,
\newblock ``Attention u-net: Learning where to look for the pancreas,''
\newblock {\em arXiv preprint arXiv:1804.03999}, 2018.

\bibitem{milletari2016v}
Fausto Milletari, Nassir Navab, and Seyed-Ahmad Ahmadi,
\newblock ``V-net: Fully convolutional neural networks for volumetric medical
  image segmentation,''
\newblock in {\em 2016 fourth international conference on 3D vision (3DV)}.
  Ieee, 2016, pp. 565--571.

\bibitem{chen2021transunet}
Jieneng Chen, Yongyi Lu, Qihang Yu, Xiangde Luo, Ehsan Adeli, Yan Wang, Le~Lu,
  Alan~L Yuille, and Yuyin Zhou,
\newblock ``Transunet: Transformers make strong encoders for medical image
  segmentation,''
\newblock {\em arXiv preprint arXiv:2102.04306}, 2021.

\bibitem{wang2021transbts}
Wenxuan Wang, Chen Chen, Meng Ding, Hong Yu, Sen Zha, and Jiangyun Li,
\newblock ``Transbts: Multimodal brain tumor segmentation using transformer,''
\newblock in {\em Medical Image Computing and Computer Assisted
  Intervention--MICCAI 2021: 24th International Conference, Strasbourg, France,
  September 27--October 1, 2021, Proceedings, Part I 24}. Springer, 2021, pp.
  109--119.

\bibitem{lu2019deepaaa}
Jen-Tang Lu, Rupert Brooks, Stefan Hahn, Jin Chen, Varun Buch, Gopal Kotecha,
  Katherine~P Andriole, Brian Ghoshhajra, Joel Pinto, Paul Vozila, et~al.,
\newblock ``Deepaaa: clinically applicable and generalizable detection of
  abdominal aortic aneurysm using deep learning,''
\newblock in {\em Medical Image Computing and Computer Assisted
  Intervention--MICCAI 2019: 22nd International Conference, Shenzhen, China,
  October 13--17, 2019, Proceedings, Part II 22}. Springer, 2019, pp. 723--731.

\bibitem{chandrashekar2020deep}
Anirudh Chandrashekar, Ashok Handa, Natesh Shivakumar, Pierfrancesco Lapolla,
  Vicente Grau, and Regent Lee,
\newblock ``A deep learning approach to automate high-resolution blood vessel
  reconstruction on computerized tomography images with or without the use of
  contrast agent,''
\newblock {\em arXiv preprint arXiv:2002.03463}, 2020.

\bibitem{ma2023deep}
Qixiang Ma, Antoine Lucas, Houda Hammami, Huazhong Shu, Adrien Kaladji, and
  Pascal Haigron,
\newblock ``Deep-learning approach to automate the segmentation of aorta in
  non-contrast cts,''
\newblock {\em Journal of Medical Imaging}, vol. 10, no. 2, pp. 024001--024001,
  2023.

\bibitem{lee2015deeply}
Chen-Yu Lee, Saining Xie, Patrick Gallagher, Zhengyou Zhang, and Zhuowen Tu,
\newblock ``Deeply-supervised nets,''
\newblock in {\em Artificial intelligence and statistics}. Pmlr, 2015, pp.
  562--570.

\bibitem{halir1998numerically}
Radim Hal{\i}r and Jan Flusser,
\newblock ``Numerically stable direct least squares fitting of ellipses,''
\newblock in {\em Proc. 6th International Conference in Central Europe on
  Computer Graphics and Visualization. WSCG}. Citeseer, 1998, vol.~98, pp.
  125--132.

\bibitem{kingma2014adam}
Diederik~P Kingma and Jimmy Ba,
\newblock ``Adam: A method for stochastic optimization,''
\newblock {\em arXiv preprint arXiv:1412.6980}, 2014.

\end{thebibliography}

\end{document}